\documentclass[letterpaper, 10 pt, conference]{ieeeconf}  
\usepackage[utf8]{inputenc} 
\usepackage{amsmath,amsfonts,amssymb} 
\usepackage{graphicx} 
\usepackage{subfigure}
\usepackage{booktabs}
\usepackage{multirow}
\usepackage{color}
\usepackage{tabularx, booktabs}
\usepackage{amsmath}
\usepackage{tikz}
\usepackage{xcolor}
\usepackage{amsmath}
\usepackage[bookmarks=false]{hyperref}
\usepackage{algorithmic}
\usepackage{algorithm}

\IEEEoverridecommandlockouts                              
\title{\LARGE \bf
Como Mensurar a Importância, Influência e a Relevância de Usuários do Twitter? Uma análise da interação dos candidatos à presidência do Brasil nas eleições de 2018.
}

\author{Ademir Cristiano Gabardo$^{1}$, Leandro Takeshi Hattori$^{1}$, Brenda Cinthya Solari Berno$^{1}$, Matheus Gutoski$^{1}$,\\ Wagner Rodrigues Ulian Agostinho$^{1}$ e Heitor Silvério Lopes$^{1}$
\thanks{$^{1}$ Universidade Tecnológica Federal do Paraná (UTFPR), Curitiba, CPGEI - Laboratório de Bioinformática e Inteligência Computacional (LABIC)}%
}

\begin{document}

\maketitle
\thispagestyle{empty}
\pagestyle{empty}

\begin{abstract}
No mundo contemporâneo, um significativo número de pessoas utilizam serviços de redes sociais para diversos fins, que incluem, mas não se limitam à comunicação, troca de mensagens e busca por informações. Uma rede social com grande popularidade no meio político é o Twitter, um serviço de \textit{microblogging} para publicação de mensagens de até 280 caracteres, chamados "tweets", onde frequentemente políticos influentes de diversos países utilizam deste meio para difundir ideias e realizar declarações públicas. Neste trabalho, foi apresentado uma análise das conexões dos candidatos à presidência da república do Brasil no ano de 2018. Utilizando as análises de redes complexas para medir a influência e relevância foi estabelecido uma métrica capaz de quantificar a importância dos usuários na rede. Como parte da análise foi utilizado um Algoritmo Memético para detectar comunidades, grupos de vértices (\textit{tweets}) fortemente conectados evidenciando agrupamentos de usuários.
\end{abstract}

\section{INTRODUÇÃO}

Nos dias atuais, as redes sociais têm grande impacto na sociedade. Principalmente pela popularização do acesso da internet, as redes sociais estão mais presentes e acessíveis, tornando-se uma importante ferramenta de comunicação e marketing digital.

Tendo em vista a grande visibilidade ao público, a adesão destas redes sociais por figuras políticas, personalidades, e corporações têm sido cada vez maior, em específico para eventos importantes como as eleições presidenciais. Consequentemente, muitos estudos têm focado na análise do impacto destas redes nas eleições \cite{Robertson2018,allcott2018trends,barclay2018political}.

Uma importante ferramenta para avaliar as interações entre as entidades de uma rede social são as redes complexas \cite{wang2015link,gonzalez2016networked}. A partir das redes complexas é possível detectar comportamentos de forma global de interações entre elementos que compõem a rede \cite{scott2017social,newman2006modularity}. Também é possível identificar a formação de comunidades, bem como detectar entidades relevantes e como interagem com as outras entidades da rede.

Criado em Março de 2006 e com mais de 500 milhões de pessoas registradas, o Twitter é uma rede com grande adesão de políticos e personalidades~\cite{jones2017social,ott2017age}.

Via de regra, \textit{tweets} são públicos, uma mensagem publicada por um usuário está disponível em seu perfil para qualquer pessoa. Características desta rede estão a interação de forma rápida e sucinta, interação em tempo real, pouco restritiva nos conteúdos publicados, permite a publicação de diversas tipos de mídias. Atrás apenas dos EUA o Brasil tem mais 41 milhões de usuários ativos no Twitter.

O objetivo deste trabalho consiste na análise da importância, influência e a relevância dos candidatos da presidência do Brasil utilizando redes complexas, a partir de dados de interações da rede social Twitter.

\section{Uma rede complexa de interações sociais no Twitter}
Neste trabalho as interações dos candidatos no Twitter foram uma rede complexa representada por meio de um grafo $G = (V, E)$, onde $G$ é um grafo, $V$ é um conjunto de vértices e $E$ é um conjunto de arestas. Um grafo é uma abstração matemática conveniente e intuitiva formada por vértices conectados por arestas~\cite{gabardo2015analise}. No grafo, os vértices podem ser usuários do Twitter, \textit{tweets}, \textit{hashtags} ou \textit{links} para mídias (por exemplo, imagens).

Existem diversas métricas para avaliar a importância de vértices em um grafo, analisando a topologia da rede, incluindo o grau, centralidade e outras medidas de \emph{ranking}. Neste trabalho especificamente foi utilizado a influência e relevância considerando os seguintes fatores:
\begin{itemize}
    \item Grau do vértice na rede - representa a importância de um usuário baseado na frequência com que este interage com os demais usuários da rede;
    \item Número de seguidores - representa a influência (capacidade de difusão) de uma informação ou opinião emitida por um usuário da rede;
    \item Número de amigos - representa a relevância de uma informação ou opinião emitida por um usuário da rede.
\end{itemize}

No Twitter, seguidores são usuários `seguindo' um usuário específico. Quando este usuário específico publica algo, todos os seus seguidores podem verão tal publicação. Enquanto `amigos' são todos os usuários que o usuário específico está seguindo\footnote{https://developer.twitter.com/en/docs/accounts-and-users/follow-search-get-users/api-reference/get-followers-list}.

As definições destes critérios são baseadas nas seguintes suposições:
\begin{itemize}
    \item O grau dos vértice reflete sua importância no grafo;
    \item Uma pessoa que possui muitos seguidores ou amigos, mas raramente interage com os usuários da rede, é menos influente que alguém que tem uma quantidade moderada de seguidores e amigos, mas que interage frequentemente na rede;
    \item Entretanto, uma pessoa que frequentemente interage com a rede, mas não tem um número significante de seguidores e amigos é menos impactante na rede.
    \item As pessoas são mais propensas a considerar informações de fontes das quais elas conheçam (por exemplo, amigos)~\cite{narangajavana2017influence}.
\end{itemize}

Desta forma, foi considerado os usuários mais importantes da rede aqueles que participam frequentemente e que tem um número significativo de seguidores ou amigos.

\section{Materiais e métodos}\label{sec:materiais-e-metodos}
\begin{figure}[!htb]
	\centering
	\includegraphics[width=0.48\textwidth]{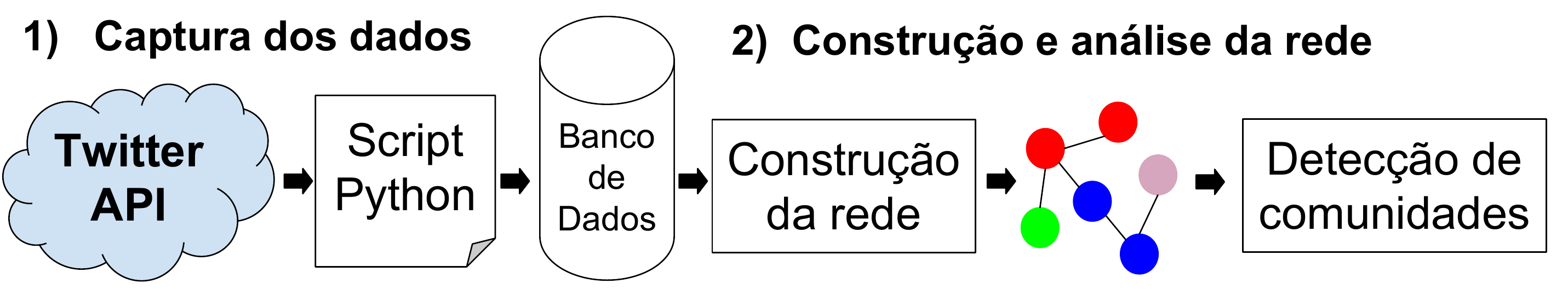} 
	\caption{Visão geral dos passos executados durante os experimentos desse trabalho.}
	\label{fig:overview}
\end{figure}

Nesta seção é detalhado o processo de captura dos dados, construção e análise da rede a partir do grupo de usuários do Twitter que compreende os candidatos à presidência da república do Brasil durante o período de campanha, conforme apresentado na Figura \ref{fig:overview}.

\subsection{Captura dos dados e construção da rede}
A API~\footnote{https://developer.twitter.com/en/docs.html} do Twitter foi utilizada para fazer a captura dos dados. O ponto de partida (\emph{seed}) são os usuários (@) dos candidatos à presidência que foram identificados como verificados como legítimos pelo Twitter\footnote{O Twitter verifica a autenticidade de usuários com significativo número de seguidores e exibe esta informação nos seus perfis}. Esta lista compreende os candidatos (usuários) listados na Tabela~\ref{tab:candidatos}.
\begin{table}[htb]
\caption{Usuários, nomes e partidos políticos dos candidatos a presidência da república do Brasil em 2018 (Em order alfabetica por nome).}
\centering
\label{tab:candidatos}
\resizebox{\columnwidth}{!}{%
\begin{tabular}{@{}lll@{}}
\toprule
Twitter Username & Nome & Partido \\ \midrule
@alvarodias\_ & Alvaro F. Dias & Podemos \\
@cabodaciolo & Benevenuto Daciolo F. S. & Avante \\
@cirogomes & Ciro Ferreira Gomes & PDT \\
@haddad\_fernando & Fernando Haddad & PT \\
@geraldoalckmin & Geraldo J. R. Alckmin Filho & PSDB \\
@guilhermeboulos & Guilherme Castro Boulos & PSOL \\
@meirelles & Henrique de Campos Meirelles & MDB \\
@jairbolsonaro & Jair Messias Bolsonaro & PSL \\
@joaoamoedonovo & João D. F. B. Amoêdo & Partido Novo \\
@joaogoulart54 & João Vicente Fontella Goulart & PPL\\
@eymaeloficial & José Maria Eymael & Democracia Cristã \\
@marinasilva & Maria O. da Silva Vaz de Lima & Rede Sustentabilidade \\
 \bottomrule
\end{tabular}
}
\end{table}

Foram coletados 5000 \textit{tweets} diariamente no período de 03/08/2018 à 29/10/2018 (61 dias). Esta quantidade de dados foi determinada arbitrariamente considerando a coleta de dados de um volume de informação suficiente para gerar uma rede significativa de interações. A Tabela~\ref{table:nodes} detalha os tipos e quantidade de vértices do grafo.
\begin{table}[htb]
\caption{Quantidade de cada tipo de interação no Twitter.}
\centering
\begin{tabular}{lr} \hline
\textbf{Tipo} & \multicolumn{1}{l}{\textbf{Quantidade}} \\ \hline
\textit{Hashtag} & 9335 \\
\textit{Link} & 7714 \\
Mídia & 15579 \\
\textit{Tweet} & 133305 \\
Usuário & 99615 \\
Total & 265548 \\ \hline
\end{tabular}
\label{table:nodes}
\end{table}

A rede complexa inicial das iterações capturadas é representada por um grafo orientado e ponderado com 265.548 vértices e 3.89.649 arestas. A Seção~\ref{subsec:communidades} detalha o método utilizado para detectar as comunidades nesta rede e a Seção~X discute os resultados encontrados.

\subsection{Encontrando comunidades em uma rede complexa}
\label{subsec:communidades}
Comunidades em redes complexas são grupos de vértices com forte relacionamento e que frequentemente exibem um alto grau de similaridade. Existem diversas métricas capazes de medir a coesão de tais grupos em redes complexas, dentre tais, modularidade é a métrica mais popular. A modularidade é uma métrica quantitativa capaz de avaliar a qualidade de uma partição de uma rede complexa. Redes com alta modularidade têm conexões densas entre os vértices de um mesmo módulo, e conexões esparsas entre vértices de comunidades distintas~\cite{newman2006modularity}. A equação~\ref{eq:modularity} mostra como calcular a modularidade $Q$ para um grafo $G$.
\begin{equation}
    Q = \frac{1}{(2m)}\sum_{vw} \left[ A_{vw} - \frac{k_v k_w}{(2m)} \right] \delta(c_{v}, c_{w}),
    \label{eq:modularity}
\end{equation}
onde $m$ representa o número de arestas no grafo, $A$ é a matriz de adjacências do grafo, $k$ indica o grau de conexão de um vértice, $vw$ representa o peso (ou valor) de uma aresta conectando o par de vétices $v$ e $w$, e $\delta$ é igual a 1 se ambos os vértices $v$ e $w$ pertencem a mesma comunidade $C$, ou zero caso contrário.

Para detectar as comunidades da rede mostrada neste estudo foi utilizado o algoritmo memético MADOC (A memetic algorithm to detect overlapping communities)~\cite{agabardoThesis,gabardo2017does}. Um algoritmo memético é um método da computação evolucionária que emprega uma população de indivíduos onde cada indivíduo representa uma possível solução a um problema. Neste caso, encontrar a partição para uma rede complexa que maximize a modularidade. Algoritmos meméticos são meta-heurísticas não determinísticas que empregam uma busca local em cooperação com a busca global de um algoritmo genético em busca de uma solução ótima ou aproximadamente ótima em um determinado espaço de busca~\cite{naeni2015ma}.


\subsection{Avaliação da importância de um vértice}
Na abordagem proposta foi avaliado a importância dos vértices sob o aspecto de topologia da rede e também sob o aspecto subjetivo ao domínio do problema. Medidas de importância de topologia da rede incluem o grau \emph{degree centrality} (que corresponde ao número de conexões de um vértice) e centralidade na rede~\cite{gabardo2015analise}.
 Na rede social Twitter, figuras públicas e instituições possuem grande quantidade de seguidores e poucos amigos (seguem menos pessoas do que seu número de seguidores), enquanto usuários comuns tendem a ter um número maior de amigos.
 Neste trabalho os seguidores foram caracterizados como entidades influentes, e o número de amigos como entidades relevantes.
 A Equação~\ref{eq:x} apresenta o cálculo da importância geral de um vértice $v(i)$ no grafo.
 \begin{equation}\label{eq:x}
    \mbox{importância de um vértice $v_i$ } = k(v_i) \times Inorm(v_i),
 \end{equation}
 onde $k(v_i)$ representa o grau (número de conexões na rede) do vértice $v_i$. O valor de $k(v_i)$ é ponderado pelo grau de importância normalizado $Inorm(v_i)$ apresentado na Equação \ref{eq:xx}.

 \begin{equation}\label{eq:xx}
    Inorm(v_i) = \frac{
    \left(
        \frac{Seguidores(v_i)}{Max(Seguidores(G))} + 
        \frac{Amigos(v_i)}{Max(Amigos(G))}
   \right)
    }{2},    
 \end{equation}
 onde $Seguidores(v_i)$ e $Amigos(v_i)$ representam a quantidade de seguidores e amigos do vértice $v_i$ respectivamente e $Max(Seguidores(G))$ (55447023 de @realdonaltrumph) e $Max(Amigos(G))$ (792755 de @FlavioBolsonaro~\footnote{incluindo os retweets}) são respectivamente a quantidade máxima de seguidores e amigos observados para um único vértice da rede no momento da coleta dos dados. Dessa forma, obtemos resultados ponderados, onde a influência e relevância dos vértices refletem sua importância na rede complexa e de sua presença online.

\section{Resultados e discução}
Nessa seção apresentamos os usuários mais importantes na rede complexa capturada, levando em conta o número de participações que reflete o grau de um vértice na rede e a métrica apresentada na Equação~\ref{eq:x}. A Tabela~\ref{table:20topDegree} mostra os 20 vértices com maior número de interações (maior grau) na rede capturada.

\begin{table}[h]
\centering
\caption{Os vinte vértices com maior número de conexões na rede}
\label{table:20topDegree}
\resizebox{0.45\textwidth}{!}{%
\begin{tabular}{llll}
\hline
Seguidores & Grau & Usuário & Nome\\
\hline
1552494 & 12148 & @jairbolsonaro & Jair Bolsonaro\\ 
700361 & 4742 & @haddad\_fernando & Fernando Haddad \\
329801 & 3135 & @cirogomes & Ciro Gomes \\
16323088 & 2094 & @danilogentili & Danilo Gentili \\
499779 & 1766 & @flaviobolsonaro & Flavio Bolsonaro \\
141222 & 1515 & @guilhermeboulos & Guilherme Boulos \\
598422 & 1513 & @bolsonarosp & Eduardo Bolsonaro \\
209440 & 1470 & @joaoamoedonovo & João Amoêdo \\
1100740 & 1368 & @geraldoalckmin & Geraldo Alckmin \\
53405 & 1168 & @buzzfeednewsbr & Buzz Feed News BR \\
409378 & 1164 & @carlosbolsonaro & Carlos Bolsonaro \\
273699 & 1012 & @oglobopolitica & O Globo Brasil \\
195981 & 989 & @joicehasselmann & Joice Hasselmann\\
51722 & 904 & @cabodaciolo & Deputado Cabo Daciolo \\
1899760 & 876 & @marinasilva & Marina Silva \\
115000 & 871 & @conexaopolitica & Blog sobre Política \\
18038 & 812 & @rejanepaiva & Rejane Paiva \\
998001 & 613 & @micheltemer & Michel Temer \\
444695 & 589 & @lulaoficial & Lula \\
14812 & 576 & @franciscomacsoa & Francisco JB \\ \hline
\end{tabular}}
\end{table}

A Tabela~\ref{table:20topDegree} é composta majoritariamente por candidatos~\footnote{O twitter @lulaoficial é a conta verificada associada ao ex-presidente Luis Inacio Lula da Silva que encontra-se preso no momento da captura dos dados, evidenciando que a conta do Twitter desse usuário é de fato gerida por outra pessoa.} e personalidades da mídia. Também é possível observar o número de seguidores para cada um dos usuários da tabela.

\begin{figure}[!h]
	\centering
	\includegraphics[width=0.4\textwidth]{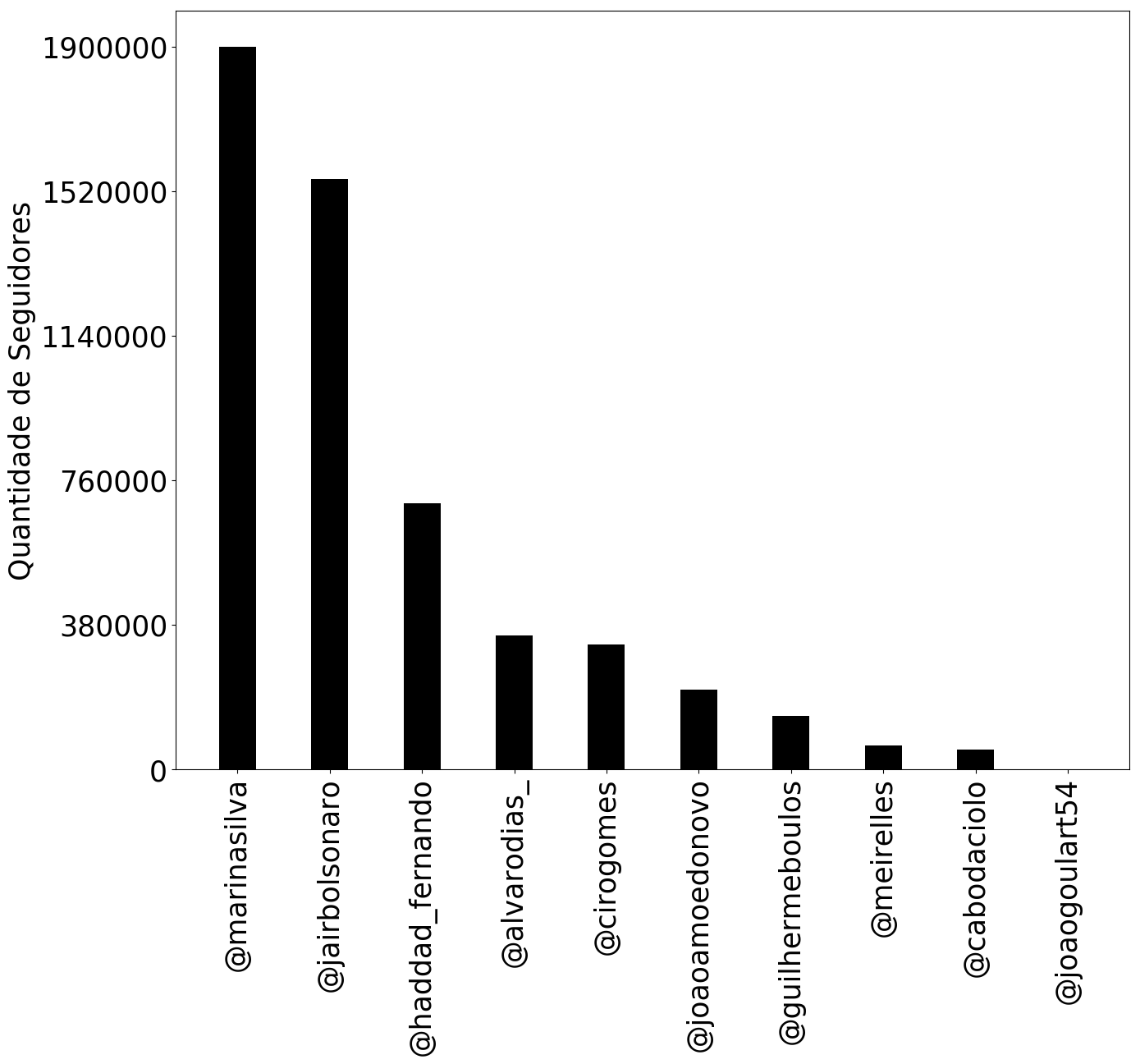} 
	\caption{Número de seguidores no twitter dos candidatos presentes na rede capturada.}
	\label{fig:seguidores}
\end{figure}

\begin{figure}[!h]
	\centering
	\includegraphics[width=0.4\textwidth]{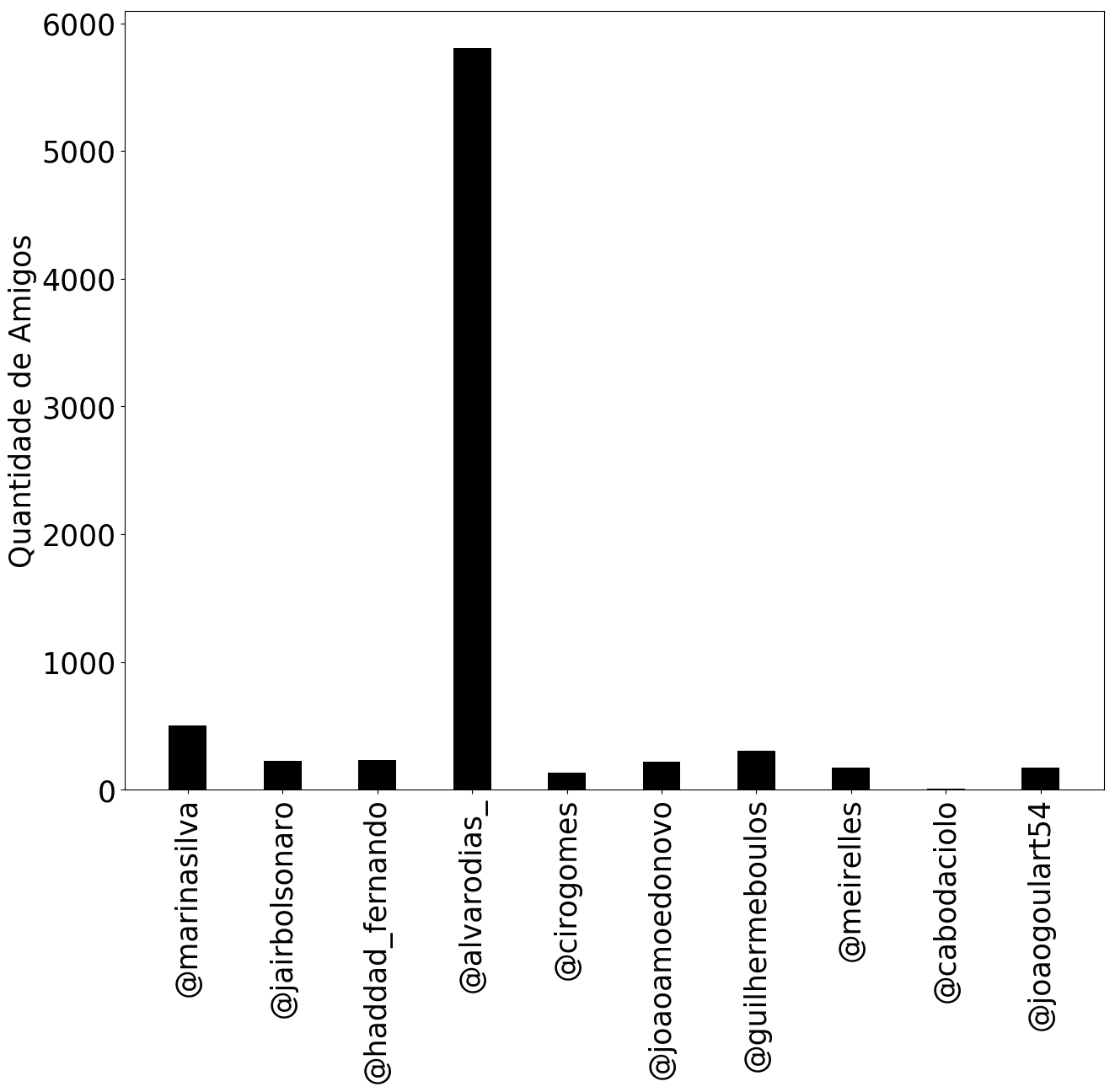} 
	\caption{Número de amigos no twitter dos candidatos presentes na rede capturada.}
	\label{fig:amigos}
\end{figure}

As Figuras~\ref{fig:seguidores} e~\ref{fig:amigos} mostram respectivamente o número de seguidores e número de amigos para cada candidato.

A Figura~\ref{fig:seguidores} mostra que a candidata com o maior número de seguidores no Twitter é Marina Silva, seguida por Bolsonaro. Na Figura~\ref{fig:amigo} observamos que o candidato Alvaro Dias segue um número consideravelmente maior de usuários no twitter, o que pode indicar o uso de \emph{Follow/Unfollow Strategy}, em que um usuário segue muitas pessoas para que estas o sigam, e mais tarde remove \emph{`unfollow'} esses usuários de sua lista, dessa forma inflando impulsionando o crescimento de sua rede de seguidores. Essa estratégia pode ser implementada com o uso de programas que vasculham a rede adicionando e removendo usuários de maneira automatizada. Esse tipo de estratégia gera seguidores, mas não engajamento~\cite{cha2010measuring}.

Eessa seção inclui também o grafo gerado a partir do componente conectado para a rede partindo dos 1000 vértices mais relevantes e suas conexões. É importante observar que alguns dos usuários do Twitter listados na Tabela~\ref{tab:importanciadegabardo} não aparecem no grafo. Por exemplo, @realdonaldtrump~\footnote{Twitter oficial do atual presidente dos EUA, não confundir @realdonaldtrump com o Twitter oficial da `entidade' presidencial dos EUA @potus \emph{45th President of the United States of America}.}  que tem um número alto de seguidores, porém nunca interagiu diretamente com nenhum dos outros usuários da lista mostrada na Tabela~\ref{tab:importanciadegabardo}, isso ocorre porque outros usuários do Twitter fizeram citações diretas ou menção a este usuário em suas postagens.

\begin{table*}[htb]
\caption{Informações dos 20 usuários mais influentes da rede capturada.}
\centering
\small
\label{tab:importanciadegabardo}
\resizebox{0.9\textwidth}{!}{%
\begin{tabular}{@{}cccclll@{}}
\toprule
Seguidores & Amigos & $k$ & Importance & Usuário & Nome & Local \\ \midrule
9570894 & 2851 & 236 & 41,161 & @marcelotas & Marcelo Tas & Sao Paulo, Brazil \\
1552494 & 228 & 495 & 13,930 & @jairbolsonaro & Jair Bolsonaro 17 & Rio de Janeiro, Brasil \\
55447023 & 46 & 12 & 12,000 & @realdonaldtrump & Donald J. Trump & Washington, DC \\
963736 & 570 & 542 & 9,615 & @micheltemer & Michel Temer & Brasília, Brasil \\
1527937 & 7273 & 254 & 8,164 & @jaarreaza & Jorge Arreaza M & Caracas, cuna de Bolívar \\
16323088 & 303 & 22 & 6,480 & @danilogentili & Danilo Gentili & Santo André \\
2458516 & 28 & 115 & 5,101 & @twittergov & Twitter Government &  \\
65988 & 30242 & 200 & 4,052 & @ddltalento & ane & @ddIevingne \\
7362380 & 314 & 26 & 3,457 & @epn & Enrique Peña Nieto & México \\
6027140 & 401 & 24 & 2,614 & @dilmabr & Dilma Rousseff & Brasília \\
8161059 & 605 & 17 & 2,.508 & @felipeneto & Felipe Neto & Rio de Janeiro \\
1821454 & 411 & 68 & 2,251 & @cdnleon & Leon Martins & Vancouver - BC \\
4381100 & 4318 & 24 & 1,961 & @mariacorinaya & María Corina Machado & Venezuela \\
18159129 & 108 & 5 & 1,637 & @10ronaldinho & Ronaldinho Gaúcho & Rio de Janeiro, Brasil \\
499779 & 1172 & 160 & 1,560 & @flaviobolsonaro & Flavio Bolsonaro 177 Senador\_RJ & Rio de Janeiro \\
2151008 & 20688 & 30 & 1,555 & @sebastianpinera & Sebastian Piñera & Chile \\
827949 & 792755 & 3 & 1,544 & @humbertotweets & Repúblico Humberto González & Caracas, Venezuela \\
42411814 & 876 & 2 & 1,530 & @nytimes & The New York Times & New York City \\
1845141 & 495210 & 4 & 1,382 & @emol & Emol.com & Santiago, Chile \\ \bottomrule
\end{tabular}
}
\end{table*}

A Tabela~\ref{tab:importanciadegabardo} também mostra apenas os dois candidatos eleitos para o segundo turno no grupo dos 20 usuários mais importantes da rede complexa desse experimento. Figuras públicas na mídia nacional (apresentadores de televisão) destacam-se na tabela por ter um grande número de seguidores no Twitter e por ter interagido diversas vezes com os usuários da rede capturada.

A Tabela~\ref{tab:importanciadegabardo} mostra ainda que usuários `comuns', que não são veículos de comunicação oficial e também não são políticos tem importante papel na difusão de informação no Twitter durante a campanha presidencial. Outro fator importante a ser observado é a heterogeneidade da localização destes usuários, incluindo pessoas e entidades fora do Brasil.

\begin{figure*}[!h]
	\centering
	\includegraphics[width=0.94\textwidth]{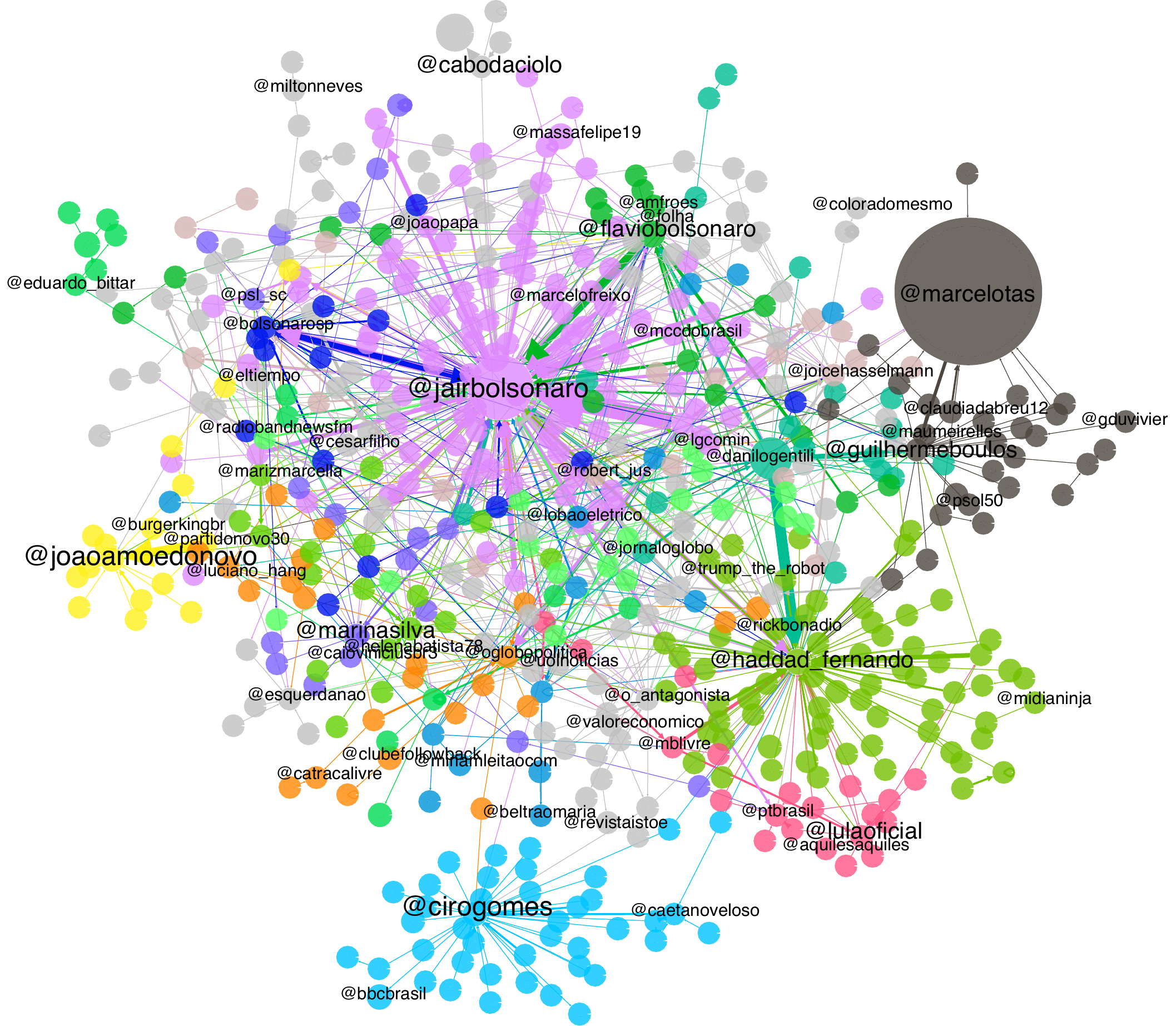} 
	\caption{Componente conectado mostrando a interação entre os usuários do Twitter com maior relevância de acordo com a importância definida na Equação~\ref{eq:x}. Dados e outras visualizações disponíveis em https://github.com/agabardo/Twitter-Brazil-Presidential-Election-2018.}
	\label{fig:primaria}
\end{figure*}

A Figura \ref{fig:primaria} apresentada o principal componente conectado, onde cada vértice possui ao menos uma ligação com o núcleo da rade, formado pelos 1000 usuários mais importantes de acordo com os critérios estabelecidos na Equação~\ref{eq:x}. Essa rede complexa foi dividida em 30 comunidades com modularidade $Q = 0.492$, as comunidades identificadas na rede são representadas na Figura~\ref{fig:primaria} por cores distintas. Os tamanhos dos vértices refletem sua importância. A direção das setas representa o fluxo de informação (origem, destino) e a espessura das arestas representa o número de vezes que uma interação se repetiu.

Na Figura~\ref{fig:primaria} podemos observar que as comunidades detectadas se formam `ao redor' de candidatos ou de figuras públicas influentes. O vértice da rede com o maior número de interações é do usuário @jairbolsonaro. Também fica evidente um considerável número de interações entre o usuário @marcelotas e o candidato do Partido dos Trabalhadores Guilherme Boulos @guilhermeboulos. Marcelo é jornalista, autor e diretor de TV e faz fortes críticas ao Partido dos trabalhadores de regimes políticos de esquerda. Diversas outras interações entre mídia e candidados ficam evidentes, como por exemplo @danilogentili e @haddad\_fernando. Danilo Gentili também é apresentador de TV que faz fortes críticas aos partidos de esquerda.

\section{Conclusões}

Neste trabalho foi aplicado análise de redes sociais e métricas de complexas para a análise da influência dos candidatos à presidência do Brasil no ano de 2018. Também propomos uma nova métrica ponderando a centralidade de grau que reflete o número de participações na rede pelo número de seguidores e amigos, desta forma, gerando uma visualização intuitíva sobre a importância dos vértices na rede.

Trabalhos futuros incluem a análise temporal dos dados capturados, essa análise já foi iniciada por outro membro do LABIC.

\section*{AGRADECIMENTOS}
Os autores agradecem à UTFPR, CPGEI, CAPES, CNPq e Fundação Araucária.

\bibliography{ref}
\bibliographystyle{IEEEtran}
\end{document}